\def\year{2021}\relax
\documentclass[letterpaper, review]{article} 

\usepackage[title]{appendix}
\usepackage{aaai21}  
\usepackage{times}  
\usepackage{helvet} 
\usepackage{courier}  
\usepackage[hyphens]{url}  
\usepackage{graphicx} 
\usepackage{graphicx}  
\usepackage{natbib}  
\usepackage{caption} 

\usepackage{comment}
\usepackage{amsmath}
\usepackage{amsfonts}
\usepackage{paralist}
\usepackage{subfigure}
\usepackage[switch]{lineno}

\newcommand{\sens}{\mathbf{s}}

\newcommand{\act}{a}

\newcommand{\agparams}{\boldsymbol{\theta}}

\newcommand{\sensstream}{S}

\newcommand{\goalstream}{G}

\newcommand{\eps}{\epsilon}

\newcommand{\advstream}{A}

\newtheorem{myLemma}{Lemma}
\newtheorem{myDefinition}{Definition}
\title{Reinforcement Learning with a Disentangled Universal \\ Value Function for Item Recommendation}

\author{

    Kai Wang, \textsuperscript{\rm 1}
    Zhene Zou, \textsuperscript{\rm 1}
    Qilin Deng, \textsuperscript{\rm 1}
    Jianrong Tao, \textsuperscript{\rm 1}
    Runze Wu,  \textsuperscript{\rm 1}\thanks{Corresponding author.} \\
    Changjie Fan, \textsuperscript{\rm 1}
    Liang Chen, \textsuperscript{\rm 2}
    Peng Cui \textsuperscript{\rm 3}
    \\
}
\affiliations{

    \textsuperscript{\rm 1}Fuxi AI Lab, NetEase Inc., Hangzhou, China\\

    \textsuperscript{\rm 2}School of Data and Computer Science, Sun Yat-sen University, China\\
    \textsuperscript{\rm 3}Tsinghua University, China\\
    \{wangkai02,zouzhene,dengqilin,hztaojianrong,wurunze1,fanchangjie\}@corp.netease.com, \\
    chenliang6@mail.sysu.edu.cn, cuip@tsinghua.edu.cn

}

\begin{document}
	
	\maketitle
	
	\begin{abstract}
		In recent years, there are great interests as well as challenges in applying reinforcement learning (RL) to recommendation systems (RS). In this paper, we summarize three key practical challenges of large-scale RL-based recommender systems: massive state and action spaces, high-variance environment, and the unspecific reward setting in recommendation. All these problems remain largely unexplored in the existing literature and make the application of RL challenging.
		
        We develop a model-based reinforcement learning framework, called GoalRec. Inspired by the ideas of world model (model-based), value function estimation (model-free), and goal-based RL, a novel disentangled universal value function designed for item recommendation is proposed. It can generalize to various goals that the recommender may have, and disentangle the stochastic environmental dynamics and high-variance reward signals accordingly. As a part of the value function, free from the sparse and high-variance reward signals, a high-capacity reward-independent world model is trained to simulate complex environmental dynamics under a certain goal. Based on the predicted environmental dynamics, the disentangled universal value function is related to the user's future trajectory instead of a monolithic state and a scalar reward.
        We demonstrate the superiority of GoalRec over previous approaches in terms of the above three practical challenges in a series of simulations and a real application.
	\end{abstract}
	
	\section{Introduction}

With the recent tremendous development of reinforcement learning (RL), there has been increasing interest in adopting RL for recommendations~\cite{shani2005mdp, hubo, ganrl}.
The RL-based recommender systems treat the recommendation process as a sequential interaction between the user  (environment) and the recommendation agent (RL agent). And a part of  user feedback (e.g., user purchase) is regarded as reward signals.
The RL-based recommender systems can achieve two key advantages: (i) the recommendation agent can explore and exploit extremely sparse user-item feedback through limited user-agent interactions; (ii) the best strategy is to maximize users' overall long-term satisfaction without sacrificing the recommendations' short-term utility.

While RL has shown considerable success in games~\cite{dqn,curious} and robotics~\cite{lillicrap2015continuous,her}, large-scale deployment of RL in real-world applications has proven challenging~\cite{dulac2019challenges}.
Compared to other machine learning methods, deep reinforcement learning has a reputation for being data-hungry and is subject to instability in its learning process~\cite{henderson2018deep}.
In this paper, we first summarize three key challenges when applying RL to RS:

\begin{itemize}
    \item \textbf{Massive state and action spaces.} In an industrial setting, the dimension of user features (state space) is extremely large, and the item is usually represented by high-dimensional item features (action space). 
    However, most model-free RL methods as well as RL-based RS methods~\cite{virtualtaobao, hubo} in the literature learn from reward signals directly and often only use small neural networks with few parameters. 
	As discussed in \citet{world}, the famous ``credit assignment problem'' and low-quality reward signals are the two main bottlenecks that make it hard for RL algorithms to learn millions of weights of a large model.

    \item \textbf{High-variance environment.} 
    Different from static gym environments, real-world recommendation environments are commonly high-variance and uncertain. On the one hand, the feedback reward of users is usually sparse (e.g., 0.2\% conversion rate) and unbalanced (e.g., a wide range of deal price). On the other hand, the random page jumps of users make the calculation of expected rewards difficult.
    These difficulties lead to a high-variance and biased estimation of the expected reward, which probably misleads the reinforcement learning towards poor performance.

    \item \textbf{Unspecific reward setting in recommendation.} Unlike reward-given environments, there is no specific reward setting given in real applications. 
    Actually, it is unclear how to do event-based reward shaping (assigning reward to user view, click, exit, etc.) to maximize business metrics (stay time, click-through rate, etc.) as well as alleviate the problem of credit assignment.
    It is common that several agents with different reward settings are deployed online simultaneously. The RL agent should be able to generalize over different reward settings and able to learn from the experiences generated by other policies, even the experiences generated by myopic supervised based recommenders.
    
\end{itemize}

In this paper, we propose a novel reinforcement learning framework called GoalRec, which takes special care of the above mentioned three practice difficulties with a disentangled universal value function:

\begin{itemize}
    \item \textbf{Handling massive state and action spaces.} Complex real applications call for specially-designed RL models which are data-efficiency and high-capacity. Motivated by the observation that the rich and temporally dense multidimensional supervision is available in recommender systems, we resort to a recently developed model-based RL technique, World Model~\cite{world, feinberg2018model}. World model is self-supervised, high-capacity, irrelevant to environmental rewards, and only predicts the environmental dynamics. 
    Different from previous pixel-based world models or one-step world model methods which only considers the next state, our RS-specific world model is designed to reconstruct the  ``measurement'' of user's long-term future trajectories under a certain goal g. Outside of the world model domain, our goal-based policy-independent world model can be seen as a generalization of long-term policy-dependent predictor \cite{tang2019disentangling, ke2019learning}.

    \item \textbf{Handling high-variance environment.} Previous value-based RL algorithms~\cite{schulman2017proximal, mnih2016asynchronous} directly estimate the value function but ignoring the coupling of environmental dynamics and reward signals.
     In contrast to the coupling manner, several works \cite{kulkarni2016deep, tang2019disentangling} are developed to decouple state transitions and reward functions for stabilizing the value estimation process by combining model-based learning and model-free value function estimation.
     In this paper, we borrow the idea of decoupling of these works and further extend it to goal-based RL. Specifically, we incorporate the powerful RS-specific  world model (model-based) into universal value function estimation (model-free and goal-based).

    \item \textbf{Handling unspecific reward setting.} 
    The problem that different reward settings with the same environmental dynamics is well studied in the goal-based RL domain.
    In this paper, we borrow the idea of goal-based RL by dealing with a more general setting of vectorial rewards and parameterized goals.
    The policy implied in users' trajectories are represented by goal vectors, and the environmental reward can be defined as the ``distance'' between current state and agent's desired goal.
    Specifically, we extend the disentangled value function to both states and goals by using universal value function (UVF).
    By universal, it means that the value function can learn from the experiences generated by other goals, and generalize to any goal g that the recommender may have. To the best of our knowledge, it is the first attempt to think multi-step RS problems in the lens of goal-based RL.
\end{itemize}

We note that GoalRec is a specially-designed unified solution rather than a trivial combination of existing developed techniques.
The decoupling of stochastic environmental dynamics and low-quality reward signals makes the training of high-capacity RL models possible. 
As the cornerstone of the whole algorithm, the world model, which is trained using temporally dense supervisory signals, can effectively alleviate the optimization issues caused by the problems mentioned above.
Instead of employing time-consuming model predictive control~\cite{hafner2018learning,ke2019learning}, we then incorporate world model into value function estimation.
Recall that the vanilla value function is policy-dependent and concerned with the expected cumulative reward over trajectories, that is why the goal-based RL and trajectory-based world model are technically necessary.

\section{Background}
\subsection{Reinforcement Learning}
The essential underlying model of reinforcement learning is Markov Decision Process (MDP). An MDP is defined as $\left< \mathcal{S}, \mathcal{A}, \mathcal{P}, \mathcal{R}, \gamma \right>$. $\mathcal{S}$ is the state space. $\mathcal{A}$ is the action space.  $\mathcal{P}: \mathcal{S} \times \mathcal{A} \times \mathcal{S} \mapsto [0,1]$ is the state transition function. $\mathcal{R}: \mathcal{S} \times \mathcal{A} \mapsto \mathbb{R}$ is the reward function. $\gamma \in [0,1]$ is the discount rate. The objective of an agent in an MDP is to find an optimal policy \(\pi_{\theta}: \mathcal{S} \times \mathcal{A} \mapsto[0,1]\) which
maximizes the expected cumulative rewards from any state \(s \in \mathcal{S}\), i.e., \(V^{*}(s)=\max _{\pi_{\theta}} \mathbb{E}_{\pi_{\theta}}\left\{\sum_{k=0}^{\infty} \gamma^{k} r_{t+k} \mid s_{t}=s\right\}\). Here \(\mathbb{E}_{\pi_{\theta}}\) is the
expectation under policy \(\pi_{\theta}\), $t$ is the current timestep and $r_{t+k}$
is the immediate reward at a future timestep $t+k$.

\subsection{Universal Value Function Approximation}

Consider for example the case where the agent's goal is described by a single desired state: it is clear that there is just as much similarity between the value of nearby goals as there is between the value of nearby states. 
A sufficiently expressive function approximator can in principle identify and exploit structure across both s and g. By universal, it means that the value function can generalize to any goal g in a set G of possible goals.

Specifically, for any goal $\mathrm{g} \in \mathcal{G}$, we define a pseudo-reward function $R_{\mathrm{g}}\left(s, a, s^{\prime}\right)$ and a pseudo-discount function $\gamma_{\mathrm{g}}(s)$. 
For any policy $\pi : \mathcal{S} \mapsto \mathcal{A}$ and each $\mathrm{g}$, and under some technical regularity conditions, we define a general state-action value function that represents the expected cumulative pseudo-discounted future pseudo-return
where the actions are generated according to $\pi$:

\begin{equation}
Q_{\mathrm{g}, \pi}(s, a) :=\mathbb{E}_{s^{\prime}}\left[R_{\mathrm{g}}\left(s, a, s^{\prime}\right)+\gamma_{\mathrm{g}}\left(s^{\prime}\right) \cdot V_{\mathrm{g}, \pi}\left(s^{\prime}\right)\right]
\end{equation}

\subsection{Decoupled Value Function}
The idea of combining model-based learning and model-free value function estimation, here we termed as decoupled value function, are widely explored in Successor Feature~\cite{kulkarni2016deep} and value decomposed DDPG with future prediction (VDFP) \cite{tang2019disentangling}. 
In this paper, we employ a similar decoupled value function formalization of VDFP.
Here, we quote the definition and corresponding lemma of VDFP to provide a brief introduction of decoupled value function.
Given a trajectory \(\tau_{t: t+k}=\left(s_{t}, a_{t}, \ldots, s_{t+k}, a_{t+k}\right)\), we consider a representation function $f$ that \(m_{t: t+k}=f\left(\tau_{t: t+k}\right)\):

\begin{myDefinition}
\label{definition:P}
Given the representation function $f$, the predictive dynamics function $P$ denotes the expected representation of the future trajectory for performing action $a \in \mathcal{A}$ in state $s \in \mathcal{S}$, then following a policy $\pi$:
\begin{equation}
\label{eqation:4}
\begin{aligned}
  P^{\pi}( s, a ) &= \mathbb{E}\left[f\left(\tau_{0: T}\right) \mid s_{0}=s, a_{0}=a ; \pi\right]\\
  &= \mathbb{E} \left[m_{0: T} \mid s_{0}=s, a_{0}=a ; \pi\right].
\end{aligned}
\end{equation}
\end{myDefinition}


\begin{myLemma}
\label{lemma:lower_bound}
Given a policy $\pi$, the following lower bound of the $Q$-function holds for all $s \in \mathcal{S}$ and $a \in \mathcal{A}$, when function $U$ is convex:

\begin{equation}
\begin{aligned}
  Q^{\pi}(s,a) &\ge U \big(f^{\pi}(s,a) \big), \\
  \mbox{where}\ \ \ \ \ \ \  
  U(\mathrm{m}) &= r_0 + \gamma r_{1} + \dots + \gamma^{t} r_{t}
\end{aligned}
\end{equation}
\end{myLemma}
The proof can be obtained with \emph{Jensen's Inequality} by 
exchanging the expectation and function, and the equality guarantees when $U$ is a linear function (the input $M^{\pi}(s,a)$ can be non-linear). 

\subsection{Related Work}
In the RL-based RS domain, past efforts mainly focused on item-list recommendation~\cite{list, page, huzhang2020validation, slateq}, simulation environment construction~\cite{ganrl, virtualtaobao, bai2019model}, variance reduction~\cite{hubo, chen2019top} and long-term reward modeling~\cite{longtermuser, drn}. Few works deal with the critical practice challenges mentioned above.
To our best knowledge, the only existing research on world model-based RS is the Pseudo Dyna-Q conducted by \citet{zou2020pseudo}, in which the static simulation environment is replaced by a constantly updated world model and the overall learning process is the same as the vanilla Dyna-Q~\cite{sutton1991dyna}. 
Not only are the problems concerned different between our work and Pseudo Dyna-Q, but also the use of world model: (i) our world model is trained using the whole trajectories of users while theirs only use users' immediate responses; (ii) our world model is a part of the value function while theirs acts as a pseudo sample provider.


\begin{figure*}
\begin{minipage}[b]{.48\linewidth}
  \centering
  \centerline{\includegraphics[width=6.0cm]{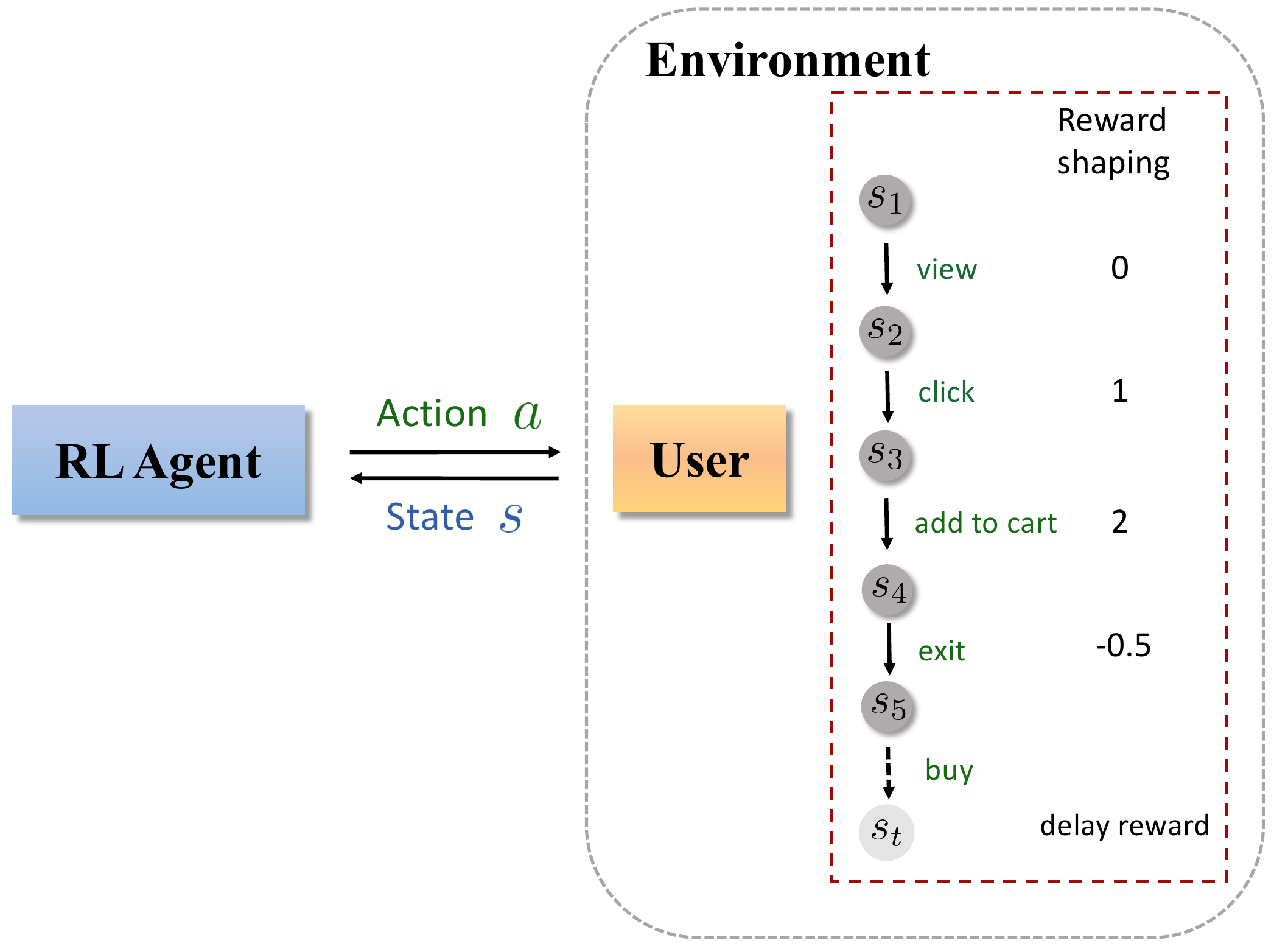}}
\centerline{(a)}
\end{minipage}
\hfill
\begin{minipage}[b]{0.48\linewidth}
  \centering
  \centerline{\includegraphics[width=6.5cm]{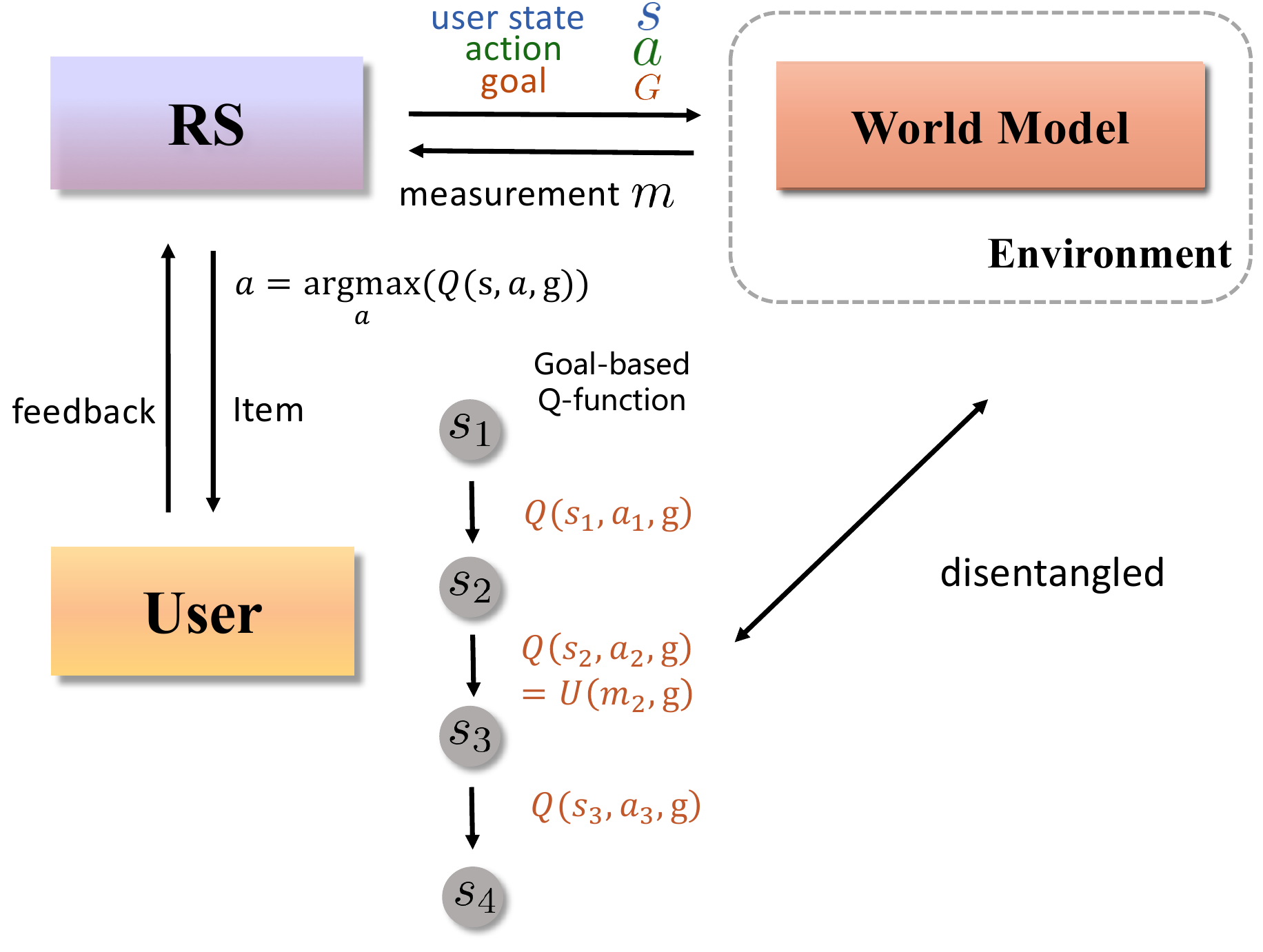}}
\centerline{(b)}
\end{minipage}
\caption{
(a) The traditional reinforcement learning framework. (b) The proposed disentangled framework. In the traditional framework, the RL agent receives the feedback of environment (user) and corresponding shaping rewards, and then train in a fully real-time interactive manner. For the GoalRec, the RS agent requests the future measurement predictor (world model) with a query consists of state $s$, action (item) $a$ and a specific goal $\mathrm{g}$, and the user is recommended for one or more items to maximize the long-term utility $Q(s, a, \mathrm{g})$.
}
\label{fig:EcUND}
\end{figure*}

\section{Proposed Approach}

In this section, we first provide an abstract description of the proposed disentangled universal value function.
Then, we discuss the specific design of  value function for recommender systems and the corresponding world model approximator--a scalable Dueling Deep Q-Network (DDQN)-like model. An illustration of the GoalRec is shown in the right part of Figure 1. The overall algorithm is summarized in Algorithm 1 in Appendix A.

\subsection{Disentangled Formalization}

Most existing world models studied in the RL community focus on pixel-based game environments and are not suitable for recommender systems. Instead of employing generative models to reconstruct the whole state space, we use the ``measurement'' of user trajectories, i.e., a measurement function $f$ defined on user trajectories $\tau$ that measurement $\mathrm{m} = f\left(s_0, a_0, ..., s_{T}, a_{T} \right)$.
Follow the decoupled value function  formalization, we have the following definitions:

\begin{myDefinition}
\label{definition:M}
The measurement $m$ of trajectory $\tau$ is a set of sufficient statistic indicators that are predictive of user response (long-term and short-term) or self-predictive (i.e., summarizes user history in a way that renders the implied environmental dynamics). 
\end{myDefinition}

\begin{myDefinition}
\label{definition:P}
Given a certain policy $\pi$ and a measurement function $f$, the function $M(s_{t}, a_{t})$ indicates the expected measurement of the partial user trajectory $\tau_{t:T}$ which starts from the step $t$ to the terminal state:
\begin{equation}
\label{eqation:4}
\begin{aligned}
   M^{\pi}( s, a ) &= \mathbb{E}_{\pi} [ \mathrm{m} | s=s_{t}, a=a_{t} ]\\
   &= \mathbb{E}_{\pi} \big[ f(\tau_{t:T}) | s=s_{t}, a=a_{t} \big].
\end{aligned}
\end{equation}
\end{myDefinition}

The measurement predictor (i.e., world model) $M$ predicts the evolution of future state and user behavior only (i.e., the environmental dynamics). It can easily incorporate the state-of-the-art offline recommendation models such as Wide\&Deep and DIEN, even the online learning algorithms like FTRL \cite{mcmahan2013ad}.
Free from the ``credit assignment problem'' and the low-quality reward signals, a high-capacity world model $M$  with millions of weights can be trained effectively to capture the complex environmental dynamics. We note that this improvement is crucial for real complex applications such as recommender systems.

Then we further extend the value function to both states and goals by using a universal value function. And the Lemma 1 still holds for a fixed goal $\mathrm{g}$.
Specifically, we replace the vanilla Q-function with a goal-based universal Q-function $Q\left(M_{\mathrm{g}}(s, a), \mathrm{g}\right)$, where $M_{\mathrm{g}}(s, a)$ is the expected measurement of user's future trajectories with the goal $\mathrm{g}$. And a linear return function $U$  that maps the expected measurement $ m=f (\tau_{t:T})$ to the aggregated discounted reward under a certain goal is defined.
\begin{myLemma}
\label{lemma:lower_bound}
Given a fixed goal $\mathrm{g}$ and the corresponding reward setting $R$, the following equality holds for all $s \in \mathcal{S}$ and $a \in \mathcal{A}$, when function $U$ is a linear function:

\begin{equation}
\begin{aligned}
  Q_{\mathrm{g}}(s,a) &= U \big(M_{\mathrm{g}}(s,a), \mathrm{g} \big), \\
  \mbox{where}\ \ \ \ \ \ \  
  U(\mathrm{m}, \mathrm{g}) &= r_0 + \gamma r_{1} + \dots + \gamma^{t} r_{t}
\end{aligned}
\end{equation}
\end{myLemma}
Different from the previous definition in which the world model $M^{\pi}$ is specific to the policy $\pi$, in the disentangled universal  value function, the world model $M_{\mathrm{g}}$ is related to the goal $\mathrm{g}$ and can generalize to different goals.
Given the composite function of $U$ and $M$ which is a strict lower-bound approximation of the $Q$-function, the best action $a^{\prime}$ in state $s$ can be selected as follows:
\begin{equation}
a^{\prime}=\max _{a} U\big(M_{\mathrm{g}}(s, a), \mathrm{g}\big)
\end{equation}




This disentangled formalization can be understood from the views of world model $M$ and return function $U$ separately. For the world model $M$, it predicts the expected future dynamics under the goal $\mathrm{g}$. Then, the linear return function estimates the expected rewards of future measurements.

\subsection{RS-specific Value Function Design }
We have described the general framework of GoalRec. In this subsection, we first introduce the definitions of state and action, measurement, and goal in recommender systems, respectively. Then, we present a RS-specific instantiation of the  goal-based Q-function. 

\textbf{State and action.} We construct four categories of features to represent state and action:
(a) Item features that describe whether certain property appears in this item, such as item's id, shop's id, brand, item's name, category, item embedding, and historical CTR in the last 1 hour, 6 hours, 24 hours and 1 week respectively.
(b) User features consist of user portrait and features of the item that the user clicked in 1 hour, 6 hours, 24 hours, and 1 week respectively. 
(c) User-item features describe the interaction between user and one certain item, i.e., the frequency for the item (also category, brand and shop) to appear in the history of the user's clicks.
(d) Context features describe the context when a request happens, including time, weekday, and the freshness of the item (the gap between the request time and item publish time).
State $s$ is represented by context features and user features.
Action $a$ is represented by item features and user-item interaction features.

\begin{figure}
	\centering
	\includegraphics[width=3.0in]{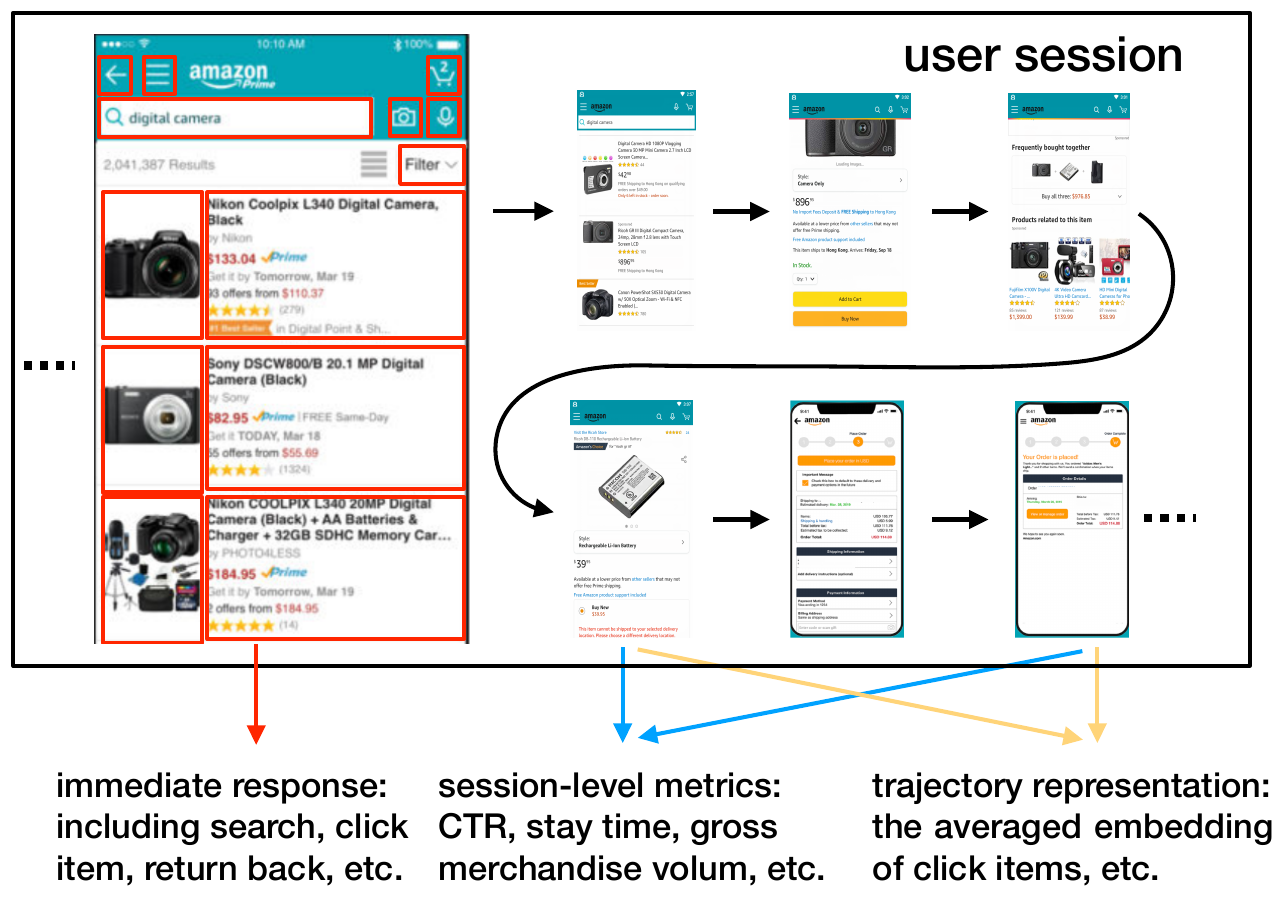}
\caption{The definition of measurement for a typical item recommender system.}
	\label{fig:measurement}
\end{figure}

\textbf{Measurement.} 
The definition of measurement is flexible and can be user-defined based on specific scenarios.
Take the typical recommender system shown in Figure~\ref{fig:measurement} as an example: the measurement can be designed as a vector composed by 
(a) rewarded user immediate responses (user purchase, etc.) and unrewarded user immediate responses (view, exit, etc.),
(b) session-level business metrics (stay time, etc.),
and (c) user's future trajectory representation (e.g., the averaged embedding of click items).
The vectorial rewards (i.e., session-level business metrics and rewarded user responses) generalize the reward setting of previous RL-based methods: the immediate user satisfaction (e.g., user purchase) or long-term user satisfaction (e.g., stay time) can be viewed as a measurement.
Moreover, the prediction of vectorial rewards, which are high-variance and sparse, can benefit from the prediction of other self-predictive measurements through the shared hidden layers.

\textbf{Goal.} In the traditional goal-based RL framework, the agent's goal is described by a single desired state, and the environmental reward can be defined as the ``distance'' between the current state and the desired goal. In our model, the goal is the desired user's future trajectories that maximize vectorial rewards. 
The goal plays two roles. First, through a distance function, the goal can be used to calculate the cumulative reward. Second, the goal can be regarded as a representation for the policies and thus determines future trajectories.

\textbf{Q-function.} Given the definitions of measurement and goal, the universal state-action value function $Q(s, a, \mathrm{g})$ can be defined as follows:
\begin{equation}
Q(s, a, \mathrm{g})=U\big(M_{\mathrm{g}}(s, a), \mathrm{g}\big)=U\big(\mathrm{m}, \mathrm{g}\big)=\mathrm{g}^{\top} \mathrm{m}
\end{equation}
where the unit-length vector $\mathrm{g}$ parameterizes the goal and has the same dimensionality as $\mathrm{m}$.

\subsection{World Model}
As discussed in the previous subsection, the problem of universal  Q-function estimation for a specific goal boils down to predicting the measurement $m$ by the world model $M$:

\begin{equation}
Q(s, a, \mathrm{g})=\mathrm{g}^{\top} \mathrm{m}=\mathrm{g}^{\top} M\left(s, a, \mathrm{g} ; \boldsymbol{\theta}\right)
\end{equation}
Here $\act \in \mathcal{A}$ is an action, $\agparams$ are the learned parameters of $M$, and $\mathrm{m}$ is the predicted future measurement. 
Note that the resulting prediction is a function of the current state $s$, the considered action $a$, and the goal $\mathrm{g}$.

\begin{figure}
	\centering
	\includegraphics[width=3.0in]{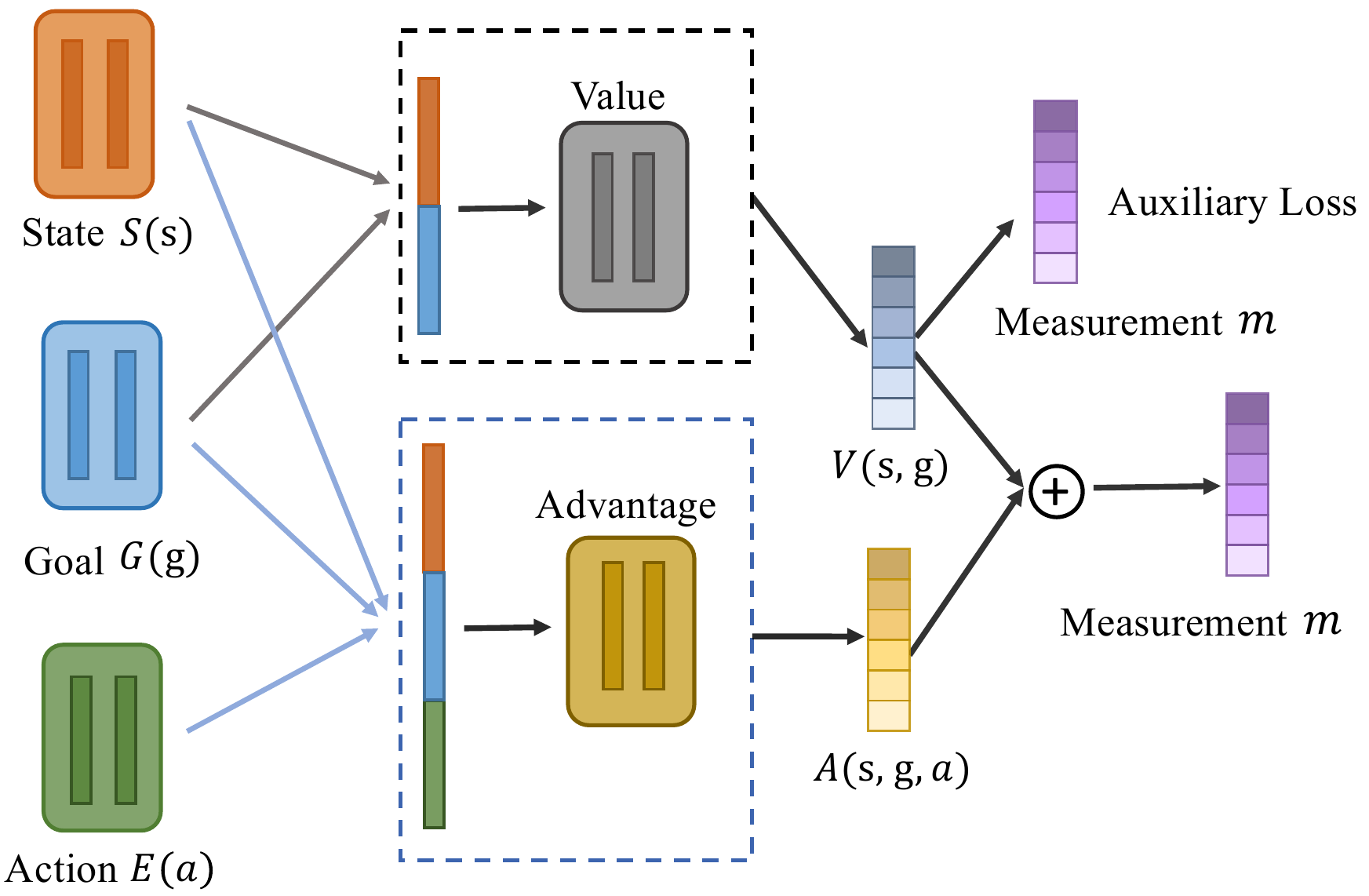}
\caption{The network structure of world model. It can be regarded as a variant of the classic two-tower recommendation model, which is scalable and able to utilize high-dimensional item features.}
	\label{fig:net_arch}
\end{figure}

The network architecture we use is shown in Figure~\ref{fig:net_arch}.
The network has three input modules: a user representation (state) module $\sensstream (\sens)$, an item representation (action) module $E(a)$ and a goal module $\goalstream (\mathrm{g})$. The state $\sens$ is represented by context features and user features, and the state module $\sensstream$ can be any state-of-the-art recommendation models such as Wide\&Deep and DIEN. The action and goal modules are fully-connected networks.

Built on the ideas of Dueling DQN~\cite{wang2015dueling}, a similar structure is employed to enhance the ability to learn the differences between the outcomes of different actions. We split the prediction module into two parts: a state value part $V(s, \mathrm{g})$ and an advantage  part $\advstream(s, \mathrm{g}, a)$.
The state value part predicts the average of future measurements over all potential actions. The action advantage part concentrates on the differences between actions. However, different from the vanilla dueling DQN, the world model does not take all actions into consideration, which means the average of the predictions over all actions is not available. To this end, we introduce an auxiliary loss to learn the state value part $V(s, \mathrm{g})$, and thus the average of the predictions of the action advantage part is nearly zero for each future measurement.
The output of these two parts has dimensionality $\dim(\mathrm{m})$, where $\mathrm{m}$ is the vector of future measurements. 
Finally, the output of the network is a prediction of future measurements for action $a$, composed by summing the output of the state value part and the action-conditional output of the action advantage part.

\subsection{Hindsight Experience Replay}
We employ a re-labeling technique, called hindsight experience replay (HER)~\cite{her}, to induce the corresponding parameterized goals (i.e., reward settings) given any trajectories. 
HER is widely used in goal-based RL, especially on the situation when the exploration of different goals is impossible (e.g., offline datasets) or costful (e.g., real applications).
Its key insight is that the agent can transform failed trajectories with no reward into successful ones by assuming that a state it saw in the trajectory was the actual goal.
In other words, since the goal does not affect the dynamics of the environment, we can re-label the trajectory afterwards to achieve different goals and thus improve sample utilization.
 For every episode the agent experiences, we store it in the replay buffer twice: once with the original goal pursued in the episode and once with the goal corresponding to the final measurement achieved in the episode, as if the agent intended on reaching this goal from the very beginning. We refer to \citet{her} for the full details of hindsight experience replay.

\subsection{Training}
Consider a set of samples collected by the agent, yielding a set $\mathcal{D}$ of training trajectories.
We then generate a training mini-batch $\mathcal{B}=\left\{\left\langle s_{i}, a_{i}, \mathrm{g}_{i}, \mathrm{m}_{i}\right\rangle\right\}_{i=1}^{N}$ from the set $\mathcal{D}$ (the steps 5 to 14 of Algorithm 1 shown in Appendix A). 
Here $\left\langle s_{i}, a_{i}, \mathrm{g}_{i}\right\rangle$ is the input and $\mathrm{m}_{i}$ is the output. 
Instead of manually tuning hyper-parameters to balance the importance of each measurement, we employ homoscedastic uncertainty trick~\cite{kendall2017uncertainties} to tune the loss weights adaptively during training.
Following the work of~\cite{kendall2018multi}, the regression loss is defined as:
\begin{equation}
\label{eq:loss}
\mathcal{L}(\boldsymbol{\theta})=\sum_{i=1}^{N} \sum_{j}^{} \left(\frac{1}{\sigma_{j}^{2}}\left\|M_{j}\left(\mathrm{s}_{i}, a_{i}, \mathrm{g}_{i} ; \boldsymbol{\theta}\right)-\mathrm{m}_{i}^{j}\right\|^{2}+\log \sigma_{j}^{2}\right)
\end{equation}
where the $j$ indicates the $j$-th measurement and the $\sigma_{j}^{2}$ is the variance of estimated Gaussian distribution of the $j$-th measurement. When $\sigma_{j}^{2}$ increases, the corresponding weight decreases. Additionally, $\log\sigma_{j}^{2}$ serves as a regularizer to avoid overfitting.
More complex loss functions such as  classification loss or reconstruction loss can be used for predicting categorical or high-dimensional measurements, but are not necessary for our experiments.
The agent follows an $\eps$-greedy policy: it acts greedily according to the current goal with probability $1-\eps$, and selects a random goal with probability $\eps$. 
The value of $\eps$ is initially set to $1$ and is decreased during training according to a fixed schedule.
The parameters of the predictor used by the agent are updated after every $N$ new experiences.

\section{Offline Experiments}
Following the setting of ~\citet{virtualtaobao} and ~\citet{ganrl}, we demonstrate how the proposed method would perform on a real-world recommender system by constructing a simulated user model on public recommendation datasets.
We empirically evaluate GoalRec to address the following questions:

Q1. Is it appropriate to apply the GoalRec on multi-step recommendation problems? How does it compare with the existing supervised learning or reinforcement learning methods?

Q2. Is the  GoalRec capable of tackling the high-dimensional state and action spaces problem in recommender systems? How about the high-variance environment?

Q3. Is the  GoalRec capable of learning a promising policy from the offline dataset generated by other policies (e.g., a myopic recommendation model)?

\subsection{Experiment Settings}

\textbf{Baselines.} 
We compare methods ranging from Supervised Learning, Model-free RL to Model-based RL, including:
\begin{itemize}
    \item \textbf{Wide\&Deep}~\cite{cheng2016wide}: a widely used state-of-the-art deep learning model combining a logistic regression and a deep neural network to predict the click label.
    
    \item \textbf{DIEN}~\cite{zhou2019deep}: another widely used state-of-the-art deep learning model which employs the attention mechanism and takes user sequence features as input.

    \item \textbf{Rainbow}~\cite{hessel2018rainbow}:  a popular off-policy value approximation paradigm, which combines several significant improvements of the DQN algorithm.
    
    \item \textbf{TD3}~\cite{fujimoto2018addressing}: an off-policy actor-critic method. The action space of TD3 is continuous. For inference, the continuous action finds its nearest neighbors in the actual item space.
    
    \item \textbf{Deep Dyna-Q}~\cite{peng2018deep}: a representative model-based approach combining one-step deep Q-learning and k-step Q-planning. 
    
    \item \textbf{GoalRec}: the vanilla GoalRec which is training from scratch and uses a goal that maximizes the cumulative reward (number of clicks) for a fair comparison. The measurement used for different datasets are listed in Appendix B. Roughly, it is a vector composed of user click/purchase/exit action, the amount/ratio of different actions, and averaged embedding of items. The state module is implemented with open-sourced DIEN algorithm\renewcommand{\thefootnote}{$4$}\footnote{https://github.com/mouna99/dien}.

    \item \textbf{GoalRec-off}: a variant that is trained on offline training datasets instead of simulation environments.

\end{itemize}

For all compared algorithms, the recommendation item list is generated by selecting the items with top-k estimated potential reward or click probability of each item. 
For a fair comparison, RL-based methods employ similar state network structures and input state features, and the item features are not considered.

\textbf{Dataset.} 
We use three real-world datasets: MovieLens-25m, Taobao and  RecSys15 YooChoose.
The dataset descriptions and detailed statistics are given in Appendix B.

\textbf{Simulation Setting.}
Following the work of ~\citet{zou2020pseudo}, we regard ``\textit{positive rating}'' or ``\textit{transactions}''  as positive feedback (clicks), and conduct the simulation with data collected from the public recommendation dataset to fit a RNN-based user model~\cite{jannach2017recurrent}. 
Apart from the feedback, the user model also predicts the users' exit probability, i.e., when to end the trajectory after losing patience.
For each dataset, we randomly select 100,000 users as the simulation environment. The users are divided as 70\% for training and 30\% for testing.
For each user, the number of interaction trajectories used for training is limited to 30 to simulate the data sparsity of real applications.
We also set up two variants of the simulation environment to evaluate algorithms in the high-dimensional state/action environment and the high-variance environment, respectively. For the high-dimensional environment, the action dimension is increased to 10,000 with the state dimension expanded to 30,000 through the discretization and intersection of features. For high-variance environment, we delayed reward signals 3 steps and add a stochastic trend noise $y_{t}=y_{t-1}+y_{t-2}+\epsilon_{t}$, where $\epsilon_{t}$ is white noise with variance $\sigma^{2}= 0.2$, on the output click probability of the estimated user model.

\textbf{Evaluation Metrics.}
The performances are evaluated by two metrics: (a) Cumulative reward:  we calculate the cumulative reward by averaging the number of clicks over all users. 
(b) CTR: the ratio of the number of clicks and the average browsing depth of users.

\textbf{Parameter Setting.}
Grid search is applied to find the optimal settings for all methods, and we report the result of each method with its optimal hyper-parameter settings. The details of grid search reported in Appendix C. 
The exploration factor $\eps$ decays from 1 to 0 during the training.
We used TensorFlow to implement the pipelines and trained networks with an Nvidia GTX 1080 ti GPU card. 

\begin{table*}[htb!]
\small
\begin{center}
    
\label{tb:policy_compare2}
\begin{tabular}{l|ll|ll|ll}
\hline
\multicolumn{1}{c}{$ $}&\multicolumn{2}{c|}{(1) MovieLens}&\multicolumn{2}{c|}{(2) Taobao}&\multicolumn{2}{c}{(3) YooChoose}\\
 \hline
model  & \multicolumn{1}{c}{Reward} &\multicolumn{1}{c|}{{ CTR}}&\multicolumn{1}{c}{Reward} &\multicolumn{1}{c|}{{ CTR}} &\multicolumn{1}{c}{Reward} &\multicolumn{1}{c}{{ CTR}}
\\ \hline
{W\&D} &  25.2($\pm$0.31) & \bf 0.601($\pm$0.006)& 2.72($\pm$0.03)& \bf 0.024($\pm$0.001) & 3.52($\pm$0.04)& 0.040($\pm$0.001)\\
{DIEN} & 26.4($\pm$0.27) & 0.592($\pm$0.007)& 2.87($\pm$0.04)& \bf 0.024($\pm$0.001) & 3.85($\pm$0.05)& \bf 0.041($\pm$0.001)\\
{Dyna-Q} & 29.3($\pm$0.80)  & 0.526($\pm$0.016) & 3.05($\pm$0.10) & 0.019($\pm$0.002) & 3.93($\pm$0.13)& 0.032($\pm$0.003)\\
{Rainbow} &  {30.1}($\pm$0.67) & {0.548}($\pm$0.012) & {3.02}($\pm$0.05) & { 0.020}($\pm$0.002) & 4.14($\pm$0.09)& 0.037($\pm$0.002)\\
{TD3} &  {28.3}($\pm$1.05) & {0.513}($\pm$0.019) & {2.80}($\pm$0.10) & {0.020}($\pm$0.002) & 3.91($\pm$0.14)& 0.034($\pm$0.003)\\
{GoalRec} &  \bf 31.6 ($\pm$0.50) & { 0.544}($\pm$0.009) &  3.14($\pm$0.06) & {0.021}($\pm$0.001) & \bf 4.23($\pm$0.07)& 0.037($\pm$0.002) \\
{GoalRec-off} &   29.1 ($\pm$0.37) & { 0.549}($\pm$0.005) & \bf 3.20($\pm$0.04) & {0.020}($\pm$0.001) &  4.09($\pm$0.05)& 0.038($\pm$0.001) \\
\hline
\end{tabular}
\caption{{Performance comparison  on the vanilla simulation environment.}}
\end{center}
\end{table*}

\begin{table*}[htb!]
\small
\begin{center}

\label{tb:policy_compare2}
\begin{tabular}{l|ll|ll|ll}
\hline
\multicolumn{1}{c}{$ $}&\multicolumn{2}{c|}{(1) MovieLens}&\multicolumn{2}{c|}{(2) Taobao}&\multicolumn{2}{c}{(3) YooChoose}\\
 \hline
model  & \multicolumn{1}{c}{Reward} &\multicolumn{1}{c|}{{ CTR}}&\multicolumn{1}{c}{Reward} &\multicolumn{1}{c|}{{ CTR}} &\multicolumn{1}{c}{Reward} &\multicolumn{1}{c}{{ CTR}}
\\ \hline
{Dyna-Q} & 27.1($\pm$0.74)  & 0.391($\pm$0.013) & 2.73($\pm$0.08) & 0.013($\pm$0.002) & 3.48($\pm$0.10)& 0.024($\pm$0.002)\\
{Rainbow} &  {28.2}($\pm$0.72) & {0.441}($\pm$0.015) & {2.61}($\pm$0.06) & { 0.015}($\pm$0.002) & 3.60($\pm$0.08)& 0.029($\pm$0.003)\\
{TD3} &  {16.2}($\pm$0.59) & {0.312}($\pm$0.024) & {1.51}($\pm$0.05) & {0.013}($\pm$0.002) & 1.86($\pm$0.07)& 0.021($\pm$0.003)\\
{GoalRec} &  \bf 33.8($\pm$0.58) & \bf {0.567}($\pm$0.01) & \bf 3.24($\pm$0.05) & \bf {0.022}($\pm$0.001) & \bf 4.35($\pm$0.07)& \bf 0.037($\pm$0.002)\\
\hline
\end{tabular}
\caption{{Performance comparison on the high-dimensional simulation environment.}}
\end{center}
\end{table*}

\begin{table*}[htb!]
\small
\begin{center}

\label{tb:policy_compare2}
\begin{tabular}{l|ll|ll|ll}
\hline
\multicolumn{1}{c}{$ $}&\multicolumn{2}{c|}{(1) MovieLens}&\multicolumn{2}{c|}{(2) Taobao}&\multicolumn{2}{c}{(3) YooChoose}\\
 \hline
model  & \multicolumn{1}{c}{Reward} &\multicolumn{1}{c|}{{ CTR}}&\multicolumn{1}{c}{Reward} &\multicolumn{1}{c|}{{ CTR}} &\multicolumn{1}{c}{Reward} &\multicolumn{1}{c}{{ CTR}}
\\ \hline
{Dyna-Q} & 28.4($\pm$0.81) & 0.513($\pm$0.015) & 2.82($\pm$0.08) & 0.014($\pm$0.002) & 3.85($\pm$0.09)& 0.023($\pm$0.002)\\
{Rainbow} &  {29.5}($\pm$0.73) & {0.530}($\pm$0.018) & {2.74}($\pm$0.06) & { 0.019}($\pm$0.001) & 3.92($\pm$0.05)& 0.032($\pm$0.002)\\
{TD3} &  {27.4}($\pm$0.95) & {0.491}($\pm$0.026) & {2.24}($\pm$0.08) & {0.016}($\pm$0.002) & 3.14($\pm$0.07)& 0.028($\pm$0.003)\\
{GoalRec} &  \bf 32.1($\pm$0.57) & \bf{0.536}($\pm$0.013) & \bf 3.02($\pm$0.05) & \bf {0.020}($\pm$0.001) & \bf 4.05($\pm$0.05)& \bf 0.034($\pm$0.003)\\
\hline
\end{tabular}
\caption{{Performance comparison on the high-variance simulation environment.}}
\end{center}
\end{table*}

\subsection{Experiment Results}

To address Q1,  we first compare our methods with non-RL and RL-based methods, and the performance is shown in Table 1. The results show that all non-RL methods, compared with DRL methods, are stable but at a lower performance level in terms of reward.
This is because they mainly focus on the item-level performance (a higher CTR) and are unable to improve the overall performance on trajectory level.
Though only equipped with the vanilla DQN, the model-based approach (Deep Dyna-Q) achieves a higher reward than Rainbow on the Taobao dataset.
Compared with Rainbow, Deep Dyna-Q, and TD3, GoalRec further improves the cumulative reward (an averaged increase of 3.7\%, 6.1\%, 10.7\%, respectively), especially on Taobao dataset.

To address Q2, we compare the performance of different models between the vanilla environment and the high-dimensional environment. As shown in Table 1 and Table 2, TD3 works badly in the high-dimensional environment because of the increase of action space. The performance of Rainbow and Deep Dyna-Q also degrade because of the ``credit assignment'' issue caused by the high-dimensional state space. Conversely, GoalRec achieves a better performance in the high-dimensional environment because of the high-capacity world model and the goal-based dense reward signals.
For the high-variance environment, the results reported in Table 1 and Table 3 show a similar conclusion. Deep Dyna-Q, Rainbow, and TD3 achieve a worse performance compared with the vanilla environment (an averaged decrease of 4.2\%, 5.5\%, 14.3\%, respectively). The decoupled value function and the adoption of world model help in alleviating the optimization issue caused by the high-variance environment.

To address Q3, we compare the performance of GoalRec-off and GoalRec. As shown in Table 1, GoalRec-off achieves a promising performance with no interactions with the simulator. The ability of learning from offline datasets directly is important since pretraining on simulation environments
may be problematic (though it is widely used in existing literatures): (i) the offline dataset is usually generated by a myopic policy and only covers limited state-action space; (ii) the supervised simulator which responded to unseen state-action confidently may mislead the RL algorithms.

Overall, the first two experiments verified the effectiveness of GoalRec when assuming the learned user simulator as ground truth. 
The third experiment shows that GoalRec offers another solution to learn from offline user trajectories and it works well. 
GoalRec is able to learn a generalization over different goals (myopic policies, etc.) from offline datasets and be adapted to a new goal (which maximizes the number of clicks) during inference.
To overcome the shortcomings of offline evaluation, we also report online experiments with real-world commercial users in Appendix C.

\section{Conclusion and Discussion}
This paper presents GoalRec, which is motivated by trying to address challenges due to large state/action spaces, high-variance environment, and unspecific reward setting. The approach decouples the model that predicts the environment dynamics and sub-information, encoded in measurements, from the way that different measurements are then encoded into rewards for various goals that the recommender may have. 
As a potential instantiation of the disentangled goal-based Q-function, GoalRec factors the value function into the dot-product between the representation of user future trajectories and a vector of parameterized goal.
It may more reasonable to think from the perspective of trajectories rather than states: recommendation environments lack significant long-range dependence between the state. By contrast, in video game environments, the key state is extremely important for future states, such as obtaining a key or entering a new scene.
The influence of the recommender agent is subtle, and the user's preference changes gradually and slowly. 

Several directions are left open in our work, including balancing explore-exploit in the goal-based RL framework 
and exploring the applicability of our solution in item-list recommendation.

\newpage

	\bibliography{sample-bibliography}
	
	\newpage
\end{document}